\documentclass[letter]{aa}
\usepackage{graphicx}
\usepackage{txfonts}
\usepackage{natbib}
\usepackage{longtable}
\usepackage[shortlabels]{enumitem}
\usepackage{siunitx}
\usepackage{epstopdf}
\usepackage{mathrsfs}

\usepackage{hyperref}	
\usepackage{amsmath}    

\hypersetup{colorlinks=true,linkcolor=blue,citecolor=blue,filecolor=blue,urlcolor=blue}
\usepackage{enumerate}%
\newcommand{\civ}{\ifmmode {\rm C}\,\textsc{iv} \else C\,\textsc{iv}\fi}

\begin{document}

\title{The G-dwarf distribution in star-forming galaxies:\\ a tug-of-war between infall and outflow}

\author { E. Spitoni\inst{1}  \thanks {email to: spitoni@phys.au.dk}, F. Calura
  \inst{2},     
  V. Silva  Aguirre\inst{1}, \and R. Gilli  \inst{2} }
\institute{Stellar Astrophysics Centre, Department of Physics and
  Astronomy, Aarhus University, Ny Munkegade 120, DK-8000 Aarhus C,
  Denmark \and INAF - Osservatorio di Astrofisica e Scienza dello Spazio di Bologna, via Gobetti 93/3, I-40129 Bologna, Italy}

\date{Received xxxx / Accepted xxxx}

\abstract
{
In the past, the cumulative metallicity distribution function (CMDF) turned out as a useful tool to constrain the accretion history of various components of the Milky Way. 
In this Letter, by means of analytical,  leaky-box chemical evolution models (i.e. including both infall and galactic outflows) we study the CMDF of local  star-forming galaxies that follow two fundamental empirical scaling relations, namely the mass-metallicity and main sequence relations. 
 Our analysis shows that galactic winds, which are dominant mostly in low-mass systems, play a fundamental role 
in shaping this function and, in particular, in determining its steepness and curvature. We show that the CMDF of low-mass (M$_{\star}$/M$_{\odot} \le 10^{9.5}$) and high-mass (M$_{\star}$/M$_{\odot}$>10$^{10.5}$) galaxies deviate substantially from the results of a 'closed-box' model, as the evolution of the former (latter) systems is mostly dominated by outflows (infall). 
In the context of galactic downsizing, we show that 
downward-concave CMDFs (associated with systems with extremely small infall timescales and with very strong winds) are more frequent in low-mass galaxies, 
which include larger fractions of young systems and  present more substantial deviations from equilibrium between gas accretion and reprocessing (either via star formation or winds).  }

\keywords{galaxies: abundances - galaxies: evolution - galaxies: fundamental parameters - ISM: general }

\titlerunning{G-dwarf distributions in star-forming galaxies}

\authorrunning{Spitoni et al.}

\maketitle

\section{Introduction}\label{introduction}

 In its original form, the G-dwarf problem consists in the fact that a 'closed-box' chemical evolution model\footnote{The basic assumptions of the closed-box chemical evolution model are i) zero initial
  metallicity; ii) homogeneity at all times; iii) a constant stellar initial mass function and
iv) constant total mass, namely no exchange of mass with the external environment.} 
predicts too many low-metallicity Galactic stars compared to the cumulative metallicity distribution
function (CMDF) observed in the solar neighbourhood \citep{bergh1962, schmidt1963}. 
Various solutions have been proposed to this problem, including a time-varying stellar initial mass function
\citep{schmidt1963,martinelli2000}, inhomogeneous star formation \citep{searle1972}, but the 
most widely-accepted solution is perhaps the introduction of an 'infall' term in the standard chemical evolution equations,
therefore relaxing the assumption that the Galactic disc behaves as a closed box \citep[e.g.][]{tinsley1978,tinsley1980, matteucci1996, matteucci2012}.
 The main effect of  considering a slow   gas accretion with primordial composition in a chemical evolution model is  the decrease of the fraction of  low-metallicity stars compared to the predictions of the closed-box model.
Following up on this idea, the adoption of an exponential infall of gas has become a fundamental assumption 
in several chemical evolution models of various components 
of our Galaxy (i.e. \citealt{matteucci1989, chiappini1997, schoenrich2009, romano2010, matteucci2019, grisoni2018,spitoni2016,spitoni2018,
spitoni2D2019,spitoni2019, spitoni2020,spitoni2021}). 

The first extragalactic investigation of the G-dwarf problem in a large sample of spiral galaxies has been recently presented by \citet{greener2021},
in their analysis of the Mapping Nearby Galaxies at APO (MaNGA;  \citealt{bundy2015}) survey data, within the fourth SDSS data release.  
Thanks to new high-quality data, \citet{greener2021} were able to  construct the CMDFs from stellar population fitting of the integrated light for a sizeable sample of galaxies as a function of stellar mass.
Their results indicate that high-mass spirals  (M$_{\star}$/M$_{\odot}$>10$^{10}$) generally show few low-metallicity stars, presenting a CMDF substantially
different than the closed-box model, hence implying a history of gas accretion similar to the one of the Milky Way disc. 
By contrast, low-mass spirals  (M$_{\star}$/M$_{\odot}$<10$^{10}$) present larger fractions of low-metallicity stars,
showing to a lesser extent clear signs of a G-dwarf problem.
In the light of these results, their interpretation is that low-mass galaxies present 
a metallicity distribution similar to the one expected if such systems evolved as closed boxes.

One important ingredient which was ignored in such interpretation is the effect of galactic winds on the 
CMDFs of the  MaNGA galaxy sample. 
Mass- and metal-loss by means of galactic outflows is fundamental to explain the metallicity of the circumgalactic and intergalactic medium
(e.g. \citealt{renzini2014}), as well as some basic scaling relations observed both in local and distant galaxies. 
A recent work where galactic winds were found to play an important role is the theoretical study presented by
\citet[][hereafter ES20]{spitoni_MZ2020} of the chemical evolution of local star-forming galaxies.
The main assumption of this work was that local star-forming galaxies
follow two fundamental, empirical scaling relations, namely the mass-metallicity relation (MZR) and the main sequence relation (MSR), namely
the observed correlation between star formation rate and stellar mass.
The study was based on  analytical, ’leaky-box’ chemical evolution models, where each galaxy originates by means of continuous accretion of
primordial gas and can also experience  outflows. The competition between these processes is fundamental
for regulating the evolution of their metal content and their star formation history. 

Using the models of ES20, in these Letter we study for the first time the roles of infall and outflow in shaping the
CDMF in galaxies as a function of their stellar mass. 
Our simple analytical formalism enables a straightforward 
interpretation of the CMDF and is useful to illustrate how the interplay between the flows of gas,
of metals and the star formation history of galaxies determine the most basic features of this particular function.
In Sect. \ref{ES20} we will summarise the ES20 model, in Sect. \ref{results} we will present our study of the CDMF and finally, in Sect. \ref{conc} we
will draw our conclusions.

\section{A summary of the ES20 model}
\label{ES20} 
\citealt{spitoni2017} presented analytical solutions to the set of differential equations for the 
evolution of the total mass, gas mass, and metallicity in galaxies. 
In their most general form, 
the equation system includes the infall 
and outflow of gas, i. e. galaxies are allowed to evolve as  leaky-box systems.
The star formation rate (SFR) is computed by means of the \citet{schmidt1959} law and  can be expressed as 
$\psi(t)= S \cdot M_\mathrm{gas}(t)$,  
where $M_\mathrm{gas}(t)$ is the gas mass at the time $t$ and $S$ is the star formation efficiency (SFE),
For this quantity we have assumed the scaling relation proposed  by \citet{boselli2014} between the typical galaxy gas depletion
timescale and stellar content. 
The gas infall rate is expressed by the exponential law $\mathcal{I}(t) = A e^{-t/{\tau}}$,
where $\tau$ is the infall timescale.
The quantity $A$ is a constant, constrained by the total infall gas mass $M_\mathrm{inf}$ \citep{spitoni2017}. The outflow rate is proportional to the SFR of the galaxy (see \citealt{recchi2008}) and is expressed as $\mathcal{W} (t) = \lambda  \cdot \psi (t)$,
with  $\lambda$  being the wind parameter (known also as loading factor, a dimensionless quantity). 
The analytical solution of \citet{spitoni2017} for the evolution of the gas-phase metallicity $Z$, defined by the ratio between the mass of metals and the gas mass 
(i.e. $Z = M_Z/M_\mathrm{gas}$),  
is: 
\begin{eqnarray}\nonumber
Z(t)&=&  \frac{y_z  \, S\big( 1-R \big)}{ \alpha \tau-1}  \cdot\\
&& \frac{M_\mathrm{gas}(0) \, t \, \big( \alpha \tau - 1 \big)^2+ A \tau\big[t - \tau  (1  + \alpha  t)    +  \tau   e^{\alpha t -t/\tau}  \big]   }
{  A \tau  \big(e^{\alpha t -t/\tau} -1\big) + M_\mathrm{gas}(0)\big(
   \alpha \tau-1\big) }, 
\label{Znew}
 \end{eqnarray}
 in which a primordial composition (i.e. metal-free) is assumed for the infalling gas  and the metallicity of the outflow is the same as the one of the interstellar medium (ISM), i.e. $Z_{out}=Z_{ISM}$. 
In Eq. (\ref{Znew}) the parameter $\alpha$ is  defined as $\alpha \equiv \big(1 + \lambda - R\big)S$.
The quantities $y_Z$ and $R$ are  the so-called yield per stellar generation and returned mass fraction.
For both quantities $y_Z$ and $R$ we use the average values computed by \citet{vincenzo2016} at various metallicities. 
Assuming a constant \citet{salpeter1955} stellar initial mass function (IMF) and the stellar yields of \citet{romano2010}, 
as metallicity-averaged values for 
the yield and for the return mass 
fraction we use $y_Z=0.0301$ and $R=0.287$, respectively. 
The other analytical expressions for the evolution of the total mass, gas mass and stellar mass can be found in \citet{spitoni2017}.

   In ES20, by adopting the aforementioned analytical solutions,  model free parameters (i.e. the total infalling mass, the  'loading factor' or wind parameter $\lambda$ and the timescale of gas accretion $\tau$) have been determined
 by imposing that the star-forming galaxies must follow the observed   local  MZR of \citet{kewley2008}  obtained with a calibration based on photoionization models provided by \citet{KD2002}
and MSR of \citet{peng2010}. We refer the reader  to Section 3 of \citet{spitoni_MZ2020} 
for further details about the methodology used to constrain the main parameters of our model.

In a 3D space formed by SFR, oxygen abundances, and stellar mass, the ancestors of these galaxies  lie on a hyper-surface, which is in good agreement with the fundamental metallicity relation  of \citet{mannucci2010}. 
In order to obey the local scaling relations, in ES20 it was found that stronger winds ($\lambda >2$) are needed in  smaller stellar mass systems. As for the infall timescale parameter, low-mass 
and high-mass galaxies can have extended distributions, ranging from 0.1 to 10 Gyr,  with slightly larger average values in the latter systems. 
By means of the same model used by ES20 to describe the backwards evolution of local systems, 
in Sect. \ref{results} we will show how the interplay between infall and galactic outflows is key for determining the shape of the CMDF in galaxies of different masses.

\section{Model results and discussion}\label{results}

Using the analytical solutions presented in 
\citet{spitoni2017}, in Sect. \ref{result1} we will analyse the main differences between the CMDF as predicted by means of closed-box and leaky-box models. 
In Sect. \ref{results2}, we will present the CMDF for the galaxies that follow the local MZR and MSR. 
Finally, in Sect. \ref{eq_sec} we will discuss further important implications of our results, 
in particular at which stage ES20  galaxies can reach a state of equilibrium (defined by a balance between their infall, outflow and star formation history) during their lifetimes and as a function of their mass, and if this stage can leave imprints on the CMDF distributions.

\subsection{Closed-box vs. leaky box  CMDFs}\label{result1}

\begin{figure}
\begin{centering}
\includegraphics[scale=0.28]{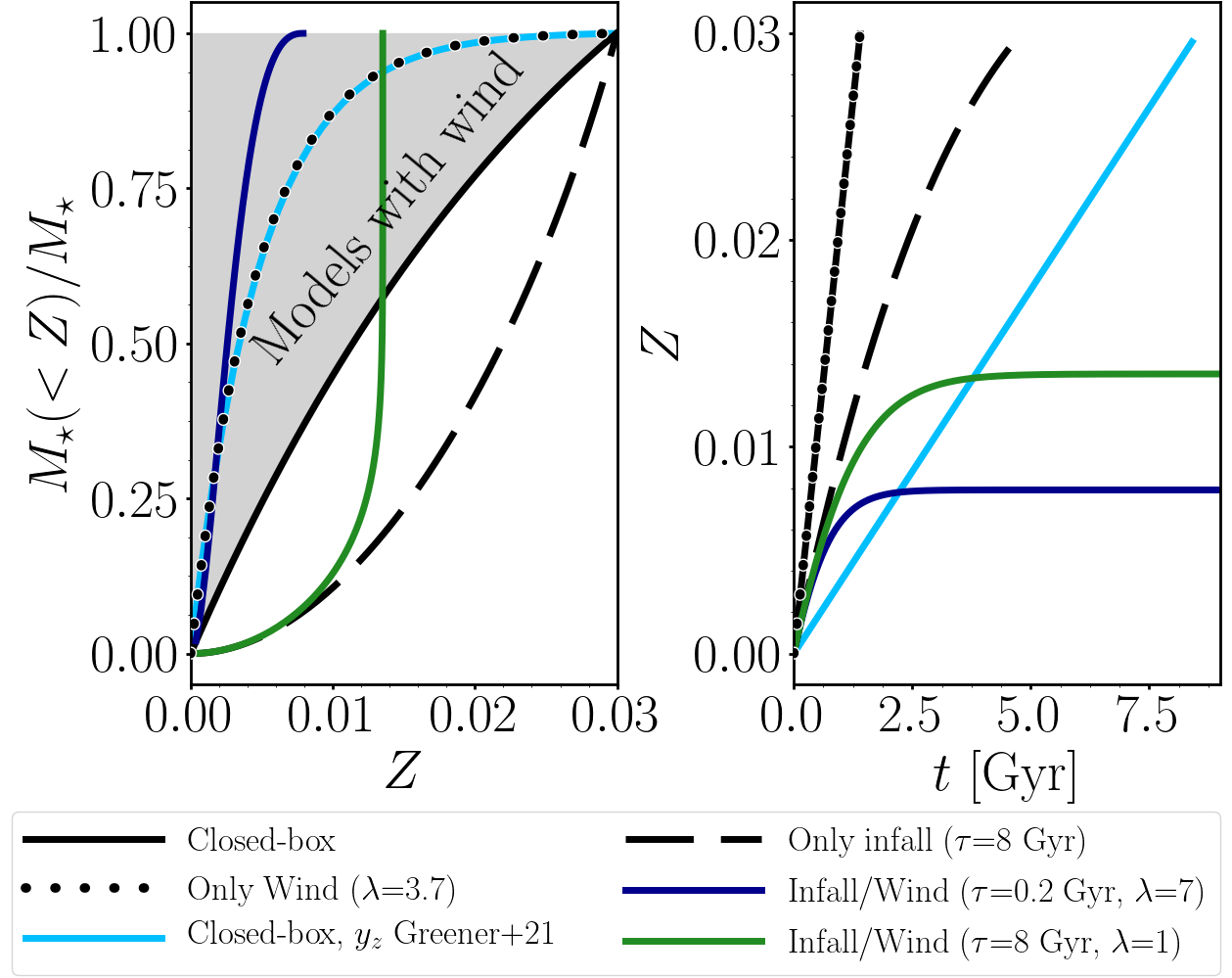}
\caption{CMDFs (left panel) and temporal evolution of the gas phase metallicity $Z$ (right panel) for  models with different prescriptions for the infall and outflow. 
The grey shaded area in the left panel indicates the region reachable only by models with outflows and assuming a yield per stellar generation $y_z=0.0301$ (as in \citealt{spitoni2017}).
The light-blue solid line shows the results of a  closed-box model obtained assuming a lower value for the yield per stellar generation  ($y_z \sim 0.005$, as in Fig. 1 of \citet{greener2021}. }
\label{4cases}
\end{centering}
\end{figure}

In Fig. \ref{4cases}, we compare the CMDF and the age-metallicity relation predicted
by a closed-box model with leaky-box systems, characterised by different values for the infall timescale $\tau$ 
and loading wind factor $\lambda$. 
In this case, the total infalling mass is fixed at the value of $M_{inf}=10^{10}$ M$_{\odot}$ and the adopted SFE is
of $S= 1$ Gyr$^{-1}$.
 However, the particular choice of these values does not affect the behaviour of the curves shown in the left panel of Fig. \ref{4cases} and their basic interpretation, as discussed in this Section. 

We consider the evolution of the closed-box (black solid line),  outflow-only (i.e.  loading factor $\lambda=3.7$ and no infall, black dotted line) and pure-infall  (i.e.  with infall parameter value $\tau=8$ Gyr and loading factor $\lambda=0$, black dashed line)
models until they reach a metallicity of $Z=0.03$, whereas their CMDFs are normalised to the total stellar mass value reached at $Z=0.03$ (as in Fig. 1 of \citealt{greener2021}). 
On the other hand, 
the other two leaky-box models with both infall and outflow reach the maximum metallicity at values smaller than $Z=0.03$. 
We notice that only the three models with a non-zero wind parameter cross the shaded area in the left panel of  Fig. \ref{4cases}. 
This basically implies that the inclusion of an outflow leads the  models to reach maximal CMDF values at lower metallicity
than the pure-infall and closed box models.  
One important feature of the curves shown in  Fig. \ref{4cases} concerns the curvature of the single tracks:
a parabolic-like behaviour (namely an upward-concave curvature) is achieved only in models with a non-zero infall parameter. 

\begin{figure*}
\begin{centering}
\includegraphics[scale=0.38]{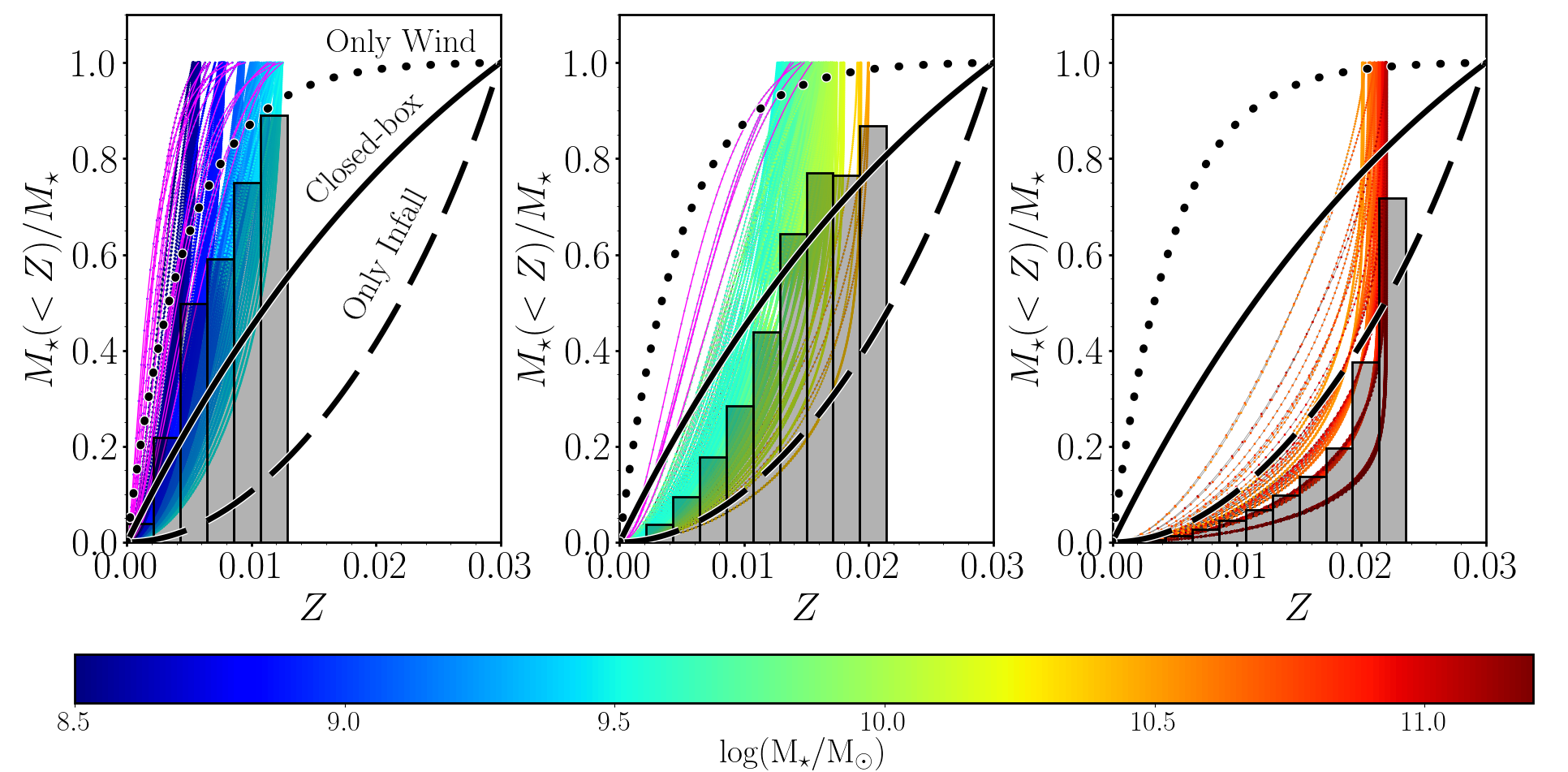}
\caption{CMDFs for star-forming galaxies which follow the local MZR and MSR (see ES20) computed in three different mass bins: 
M$_{\star}$/M$_{\odot} \le 10^{9.5}$ (left panel), 10$^{9.5}$<M$_{\star}$/M$_{\odot} \le 10^{10.5}$ (middle panel) and M$_{\star}$/M$_{\odot}$>10$^{10.5}$ (right panel). 
 In each panel, the small solid circles   connected by a thin grey line indicate the CMDFs of single 
 galaxies, colour coded according to the stellar mass at $z \sim 0.1$. The thin magenta lines are used to indicate single galaxies with downward-concave CMDFs. 
 The grey shaded histograms indicate the average values of the CMDFs computed in 14 different metallicity bins.  
The solid lines, dashed lines and dotted 
 lines represent the CMDF 
 of the closed-box model, only-infall and only-outflow model, respectively, computed with the same parameters as 
 indicated in Fig. \ref{4cases}.}  
\label{CMDF_small}
\end{centering}
\end{figure*}

Concerning the age-metallicity of the closed-box model, in \citet{spitoni2015U} it was shown that in presence of a \citet{schmidt1959} law
for the SFR ($\psi= S \cdot M_\mathrm{gas}$),  the temporal evolution of the metallicity  $Z_{cb}$ and the gas mass  can be expressed as follows:
\begin{equation}
Z_{cb}(t)=y_z (1 - R) S t \quad\mathrm{and}\quad M_{gas}(t)=M_{gas}(0)e^{-(1-R)St},
\label{acloset}
\end{equation}
respectively. These solutions can be retrieved by imposing $A=0$ (no infall) and $\lambda=0$ (no wind) in Eq. (\ref{Znew}) and
in Eq. (9) of \citet{spitoni2017}. The  linear growth of the metallicity $Z_{cb}$ as  function of time is visible in the right panel of Fig. \ref{4cases}. 
From  Eqs. (\ref{acloset}), 
we  obtain the analytical expression for the CMDF, which can be expressed as follows:
\begin{equation}
    M_{\star} \left( < Z \right)/M_{\star} = \frac{1 - e^{ -Z / y_z } }{1 - e^{ -Z_1 / y_z } }, 
	\label{eq:closed_box}
\end{equation}
where in our case  $Z_1=0.03$. The shape of the CMDF is not dependant on the particular choice of  the SFE. 
However, this is clearly not true for the evolution of the metallicity, as 
systems will reach the $Z_1$ value at different evolutionary times if  different SFE values are adopted; 
in particular the adoption of larger $S$ values will shorten the time to reach a high metallicity. 
Concerning the system with only wind ($\mathcal{W}=\lambda \cdot \psi$) and a \citet{schmidt1959} law for the SFR,  the analytical solution for the CMDF can be expressed as follows: 
\begin{equation}
    M_{\star} \left( < Z \right)/M_{\star} = \frac{ 1 - e^{ \frac{-(1-R+\lambda)Z}{(1-R)y_z}  }}{ 1 - e^{ \frac{-(1-R+\lambda)Z_1}{(1-R)y_z}}}.
	\label{eq:wind}
\end{equation}
From this expression, by imposing the no-wind condition (i.e. $\lambda=0$), it is possible to retrieve the CMDF solution for the closed-box model of Eq. \ref{eq:closed_box}.
Fig. \ref{4cases} shows that the age-metallicity for the closed-box and only outflow models are identical, in agreement with the findings of \citet{spitoni2015U}. 

An important warning is in order when comparing the CMDF of the closed-box and outflow-only models. 
We note that in Fig. 1 of \citet{greener2021} the closed-box solution is obtained assuming a 'not physically motivated' (as stated by \citealt{greener2021}) yield value of  $y_z \sim 0.005$, roughly 6 times smaller than the one adopted in our analysis.
In \citet{matteucci2012}, it was underlined that for leaky-box models in presence of outflows, the 'true yield' (defined as $y_{true}=Z/\ln [\mu^{-1}]$, where $\mu=M_{gas}/M_{tot}$) 
is always lower than the 'effective'  one obtained with closed box models  $y_{eff}=Z_{cb}/\ln (\mu_{cb}^{-1})=y_z$. 
This is basically the reason why, with a closed box model (such as the one 
\citealt{greener2021}), 
it is possible to achieve a CMDF with an identical shape to the one of our outflow-only model with $\lambda=3.7$ (see Fig. \ref{4cases}). 
In conclusion, besides discussing the roles of infall and outflow in determining the particular shape of the CMDF, we have shown a particular case in which the CMDF of a leaky-box model can be the same as the one of a closed-box model with a reduced yield. This has to be bear in mind in the interpretation of this quantity. 

 In a future work it is our intention to  take into account more sophisticated prescriptions for the gas infall/outflow rates.  
For instance, the presence of differential winds \citep{recchi2008}, namely galactic winds in which  metals  are  ejected out of the parent galaxy more efficiently than the other elements. What we expect in the case of a further depletion of metals is a saturation of the CMDF occurring at lower metallicities, i. e. a further increase of its steepness. 
Likewise, there is evidence that inflows at late times \citep{mitchell2020, voort2017} are chemically enriched ("wind recycling"). An enriched gas infall in massive galaxies  could help   alleviating the G-dwarf problem \citep{matteucci2012}.

\begin{figure}
\begin{centering}
\includegraphics[scale=0.27]{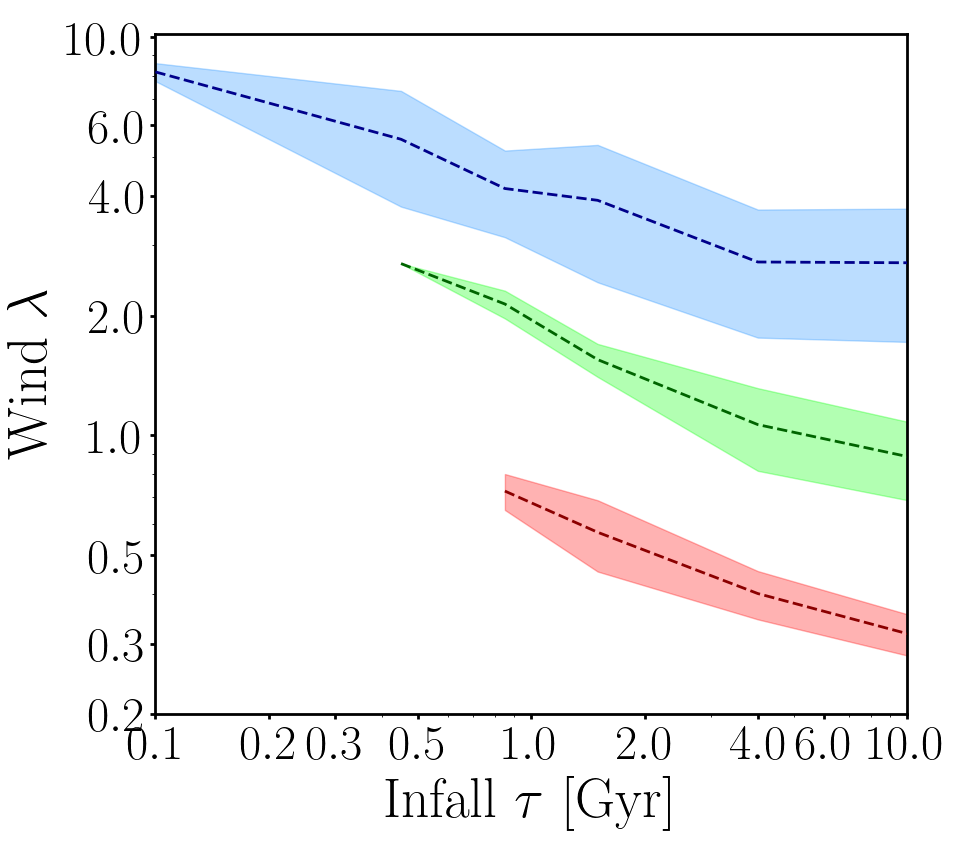}
\caption{Wind parameter $\lambda$ versus infall time scale $\tau$ of the star-forming galaxies which follow the local 
MZR and MSR (ES20) and in three 
different  stellar mass bins:
M$_{\star}$/M$_{\odot} \le 10^{9.5}$ (blue line and region), 10$^{9.5}$<M$_{\star}$/M$_{\odot} \le 10^{10.5}$ (green) and M$_{\star}$/M$_{\odot}$>10$^{10.5}$ (red). 
The dashed lines represent the average values, whereas the shaded colored regions represent the 
1$\sigma$ dispersion.}
\label{tau_lambda}
\end{centering}
\end{figure}

\begin{figure*} 
\begin{centering}
\includegraphics[scale=0.34]{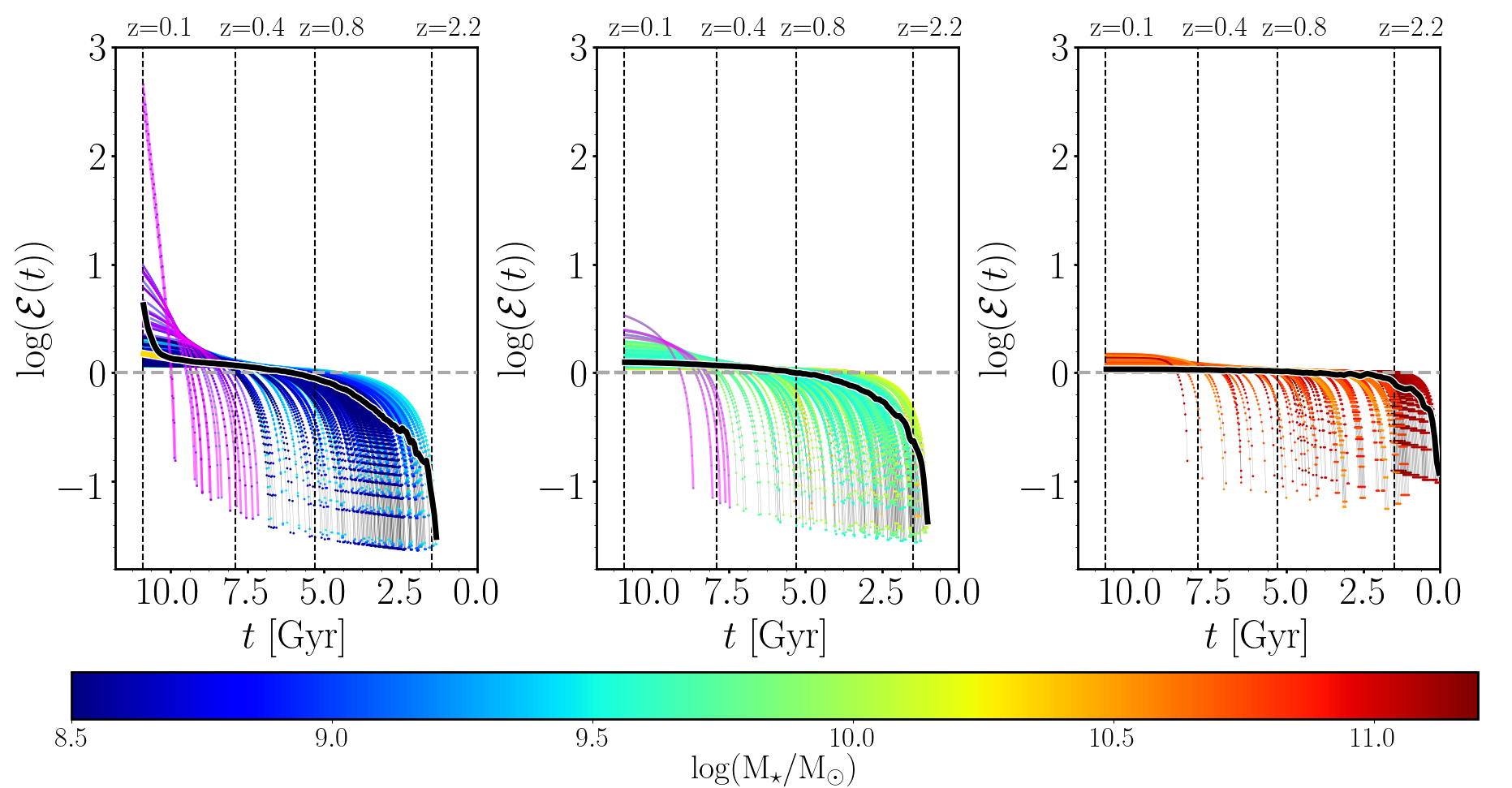}
\caption{Evolution of the quantity  $\mathcal{E}(t)$ introduced in Eq. \ref{eq_eq} as function of time for star-forming galaxies in the three stellar mass bins of Fig. \ref{CMDF_small}.
The single galaxy tracks (small solid circles connected by grey lines) are colour-coded according to their stellar mass at $z \sim 0.1$, whereas the thick black solid lines are the average  values. 
The thick yellow line in the left panel represents the average $\mathcal{E}(t)$ but computed excluding the 2 galaxies showing the largest values of the same quantity at redshift $z=0.1$. In each panel, the magenta lines represent the galaxies characterised by downward-concave  CMDFs.}  
\label{EQ}
\end{centering}
\end{figure*}

\subsection{CMDFs for star-forming galaxies of ES20}\label{results2}

In Fig. \ref{CMDF_small}, we present the CMDFs calculated with the models of ES20 in
three different stellar mass bins, namely
M$_{\star}$/M$_{\odot}\le10^{9.5}$, 10$^{9.5}$<M$_{\star}$/M$_{\odot} \le 10^{10.5}$ and M$_{\star}$/M$_{\odot}$>10$^{10.5}$. 
In each panel of Fig. \ref{CMDF_small}, the colour-coded tracks show the results obtained for each single galaxy, 
whereas the grey histogram shows the average CDMF computed in each bin. 
The results obtained by means of the ES20 model are compared to the tracks of the infall-only, outflow-only and closed box models described in Sect. \ref{result1}.  

As visible in the right panel of Fig. \ref{CMDF_small}, in the largest stellar mass bin the computed CMDFs deviate significantly from the predictions of the closed-box model, but are instead in good agreement with the infall-only model. 
 
Another noteworthy feature of the results obtained for the most massive galaxies is the complete lack of downward-concave curves, associated to systems with outflows but very short infall timescales 
(such as the  dark blue line of Fig. \ref{4cases}, with  $\tau=0.2$ Gyr).

The middle panel of Fig. \ref{CMDF_small} shows the CMDFs computed for galaxies with intermediate stellar mass.
The vast majority of galaxies in this mass bin still show an upward-concave CMDFs, but in this case we see the presence of a 
small fraction ($\sim$ 2.1\%) of models which show an opposite behaviour, i. e. downward-concave curves (as indicated by the magenta lines).
As for the average CMDF, in general it is in better agreement with the closed box model than the one of the most massive bin. 
 
To understand this result, 
one has to consider that the role of the galactic winds in the evolution of intermediate-mass galaxies is significantly 
stronger than for the most massive galaxies. 
 The fact that our galaxies are forced to 
follow the local MZR implies final metallicity 
values which are lower than the ones of the galaxies in the most massive bin. 
In general, this implies a steeper increase of the average CMDF, qualitatively more similar to the one of the closed-box model than to the ones of leaky-box models.

 An even more pronounced steepness
is evident for most of the galaxies in the lower mass bin (left panel of Fig. \ref{CMDF_small}). 
In this case, the presence of systems characterised by strong outflows and short timescales of gas accretion is 6.4 \%, hence larger than in the previous mass bin, as visible from the more significant presence of downward-concave curves. 
The systems belonging to the lowest mass bin are characterised by the lowest final metallicity values ($Z<0.01$), and
are therefore characterised by the steepest CMDFs of our sample. 
The average CMDF is in agreement with the prediction of our outflow-only model ( with $\lambda=3.7$), 
also equivalent to the closed-box model of \citealt{greener2021} with a reduced yield shown in Fig. \ref{4cases}).

In summary, the analysis of the CMDF described in  this Section confirms that the evolution of the most massive star-forming galaxies is dominated by infall, occurring on long timescales ($>1$ Gyr). 
The systems whose evolution is more similar to the one of a closed-box, which in our picture can be envisaged as characterised by a controlled balance between infall and outflows, are intermediate-mass galaxies. 
Finally, the evolution of low-mass galaxies is dominated by outflows.  
Our analysis offers a novel point of view to confirm the well-known importance of outflows in regulating the properties of galaxies, which increases as their stellar mass decreases 
\citep[e.g.][]{dekel1986,tremonti2004, hirschmann2016, lian2018ev}.

\subsection{Galactic downsizing and  equilibrium condition}\label{eq_sec}
In this Section, we discuss a few implication of 
our analysis which are mostly related to galactic downsizing. 
This is a well-known property consisting into 
substantial differences in the star formation history of low- and high-mass galaxies, with a higher star formation activity in the most massive galaxies at early epochs, followed by more intense SF in low-mass galaxies at more recent times \citep[e.g.][]{cowie1996,calura2009,caluraMZ2009, mortlock11}.

Fig. \ref{tau_lambda} is useful to focus the attention on the roles of infall and outflow in regulating the evolution of star-forming galaxies and to show the allowed range of both parameters in galaxies of different mass, as dictated by the local MZR and MSR. 
Here,  we show with different colours the regions occupied by galaxies of the same three different mass bins of Fig. \ref{CMDF_small}
in the space described by $\lambda$ and $\tau$. 
The most massive galaxies (red shaded region in \ref{tau_lambda}) occupy the bottom-right corner of the plot and are therefore characterised by the longest accretion timescales ($> 1$ Gyr) and by the smallest values of the wind parameter ($\lambda < 0.7$).
On the other hand, the largest wind parameter values ($\lambda>2$) characterise the evolution of the lowest mass galaxies (blue shaded region in Fig. \ref{tau_lambda}), which are also characterised by the most extended range of infall parameters, spanning the interval
$0.1$ Gyr $\le \tau \le$10 Gyr.

The observed correlation between galactic age and stellar mass relation is one main expression of galactic downsizing and, as shown by ES20, it stems directly from the local scaling relations of star-forming galaxies. 
In fact, at low redshift these objects have to maintain a large reservoir of gas in order to be part of the local MSR. In this regard, weak winds and long infall timescales act together against a rapid gas consumption and cease of star formation. In such conditions, a long lifetime is needed for them to be at present simultaneously metal-rich and still star-forming, which implies very old average ages and generally wide, upward-concave shapes for the CMDF.  

On the other hand, lower mass systems are allowed to form continuously, and in all of them a low star formation rate and low  metallicity, 
as required by the MZR and MSR, respectively, 
are both achieved by means of a wind parameter much stronger than more massive galaxies. 
Of these systems, those characterised by small (long) infall timescales will show a downward- (upward-) concave CMDF.

The interplay between star formation, infall and galactic winds regulates the gas content of galaxies and sets the timescale for them to reach a state of 'equilibrium', 
in which the amount of accreted matter is balanced  
by how much galactic material is processed, either via star formation or outflows  \citep{peeples2011,dayal2013,lilly2013,peng2014}. 
By means of our models, we assess at which stage of their life the star-forming galaxies of various masses can reach such equilibrium condition, which is 
defined by a null or very weak variation of the gas mass during their  evolution  \citep{maiolino2019}.

By imposing a null variation of the mass 
in the second equation of system (7) of \citet{spitoni2017}, we obtain:
\begin{equation}
\dot{M}_{gas}(t) \stackrel{\text{Equil.}}{\equiv} 0 \Rightarrow  \mathcal{E}(t) \equiv
\frac{\big(1 - R \big) \psi(t)+\lambda \psi(t)}{ A e^{-t/{\tau}}}=1. 
\label{eq_eq} 
\end{equation}
In Fig. \ref{EQ}, we show the evolution of the 'equilibrium' parameter $\mathcal{E}(t)$ defined above (i.e. ratio between reprocessed and accreted gas) in galaxies of three different mass bins as defined in Figs.  \ref{CMDF_small} and \ref{tau_lambda}. 
In this plot, values of $\mathcal{E}(t)$ close to 1 are to be associated with systems in equilibrium. 
From Fig. \ref{EQ} we can see that 
a well-defined relation between stellar mass and equilibrium timescale is in place, 
shown by the curvature of the average $\mathcal{E}(t)$ computed in the three mass bins. 
The bulk of the galaxies in the highest mass bin have started to behave as 'gas-regulators' (showing $\mathcal{E}(t)$  $\sim$1) already at redshift $z \sim 2.2$, i.e. in the earliest phases of their life. This implies that these systems have spent almost all their entire lifetime in equilibrium. 
On the other hand, on average the objects with stellar mass M$_{\star}$/M$_{\odot} \le 10^{9.5}$ are characterised by the longest equilibrium timescales.  
In all the three stellar mass bins of  Fig. \ref{EQ}, several systems show a low-redshift upturn of $\mathcal{E}$. 
This is due to the late emergence of galaxies characterised by short timescales of accretion, which
show stronger deviations from equilibrium in the lower mass bins as due to their stronger winds. 
The systems which cause the strongest deviations from
the equilibrium are those formed at the most recent times and the same characterised by downward-concave CMDF, with the smallest infall timescales and the strongest winds. 
As already discussed in Sect. \ref{results2}, the importance of this population increases as we move towards low stellar mass object, as visible also from the  average $\mathcal{E}$, significantly affected by the presence of these systems only in the stellar bin with M$_{\star}$/M$_{\odot} \le 10^{9.5}$.

 \section{Concluding remarks}\label{conc}
In this Letter, we have studied the CMDFs of  star-forming galaxies as a function of their   mass and by means of the ES20 model. 
We showed how the shape of the CMDF in galaxies is determined by a  'tug-of-war' between infall and outflow.
In agreement with the recent SDSS IV MaNGA data analysis of \citet{greener2021}, the CMDFs of the  most massive objects (M$_{\star}$/M$_{\odot}$>10$^{10.5}$) deviate substantially from the closed-box model predictions and their evolution is dominated by infall, occurring on timescales typically of a few Gyr and consistent 
with what found in the earliest studies of the G-dwarf problem in the Milky Way \citep[e.g.][]{tinsley1978,tinsley1980}.
The evolution of intermediate-mass galaxies (10$^{9.5}$<M$_{\star}$/M$_{\odot} \le 10^{10.5}$) sees a controlled balance between infall and outflow and their CDMF is similar to the one obtained with an analytic closed-box model. 
Finally, in the evolution of low-mass (M$_{\star}$/M$_{\odot} \le 10^{9.5}$) galaxies, the outflows prevail and they present a much steeper average CDMF, more similar to the one achievable with a leaky-box analytic model with extremely short infall and strong wind parameter. 
In individual galaxies, the stellar mass  is a key parameter and it determines the timescale for reaching equilibrium between inflows and processed matter (both through winds and star formation). 
The strongest deviation from such equilibrium are presented by the youngest galaxies, 
whose frequency is larger in the lowest mass bin and 
which are characterised by a peculiar, downward-concave shape of the CMDF, achievable only with extremely small infall timescales and with a very strong wind parameter.

\section*{Acknowledgement}

We thank the anonymous referee for  constructive comments, which improved the quality  of our work.
Funding for the Stellar Astrophysics Centre is provided by The Danish National Research Foundation (Grant agreement no.: DNRF106).
E. Spitoni and V. Silva Aguirre acknowledge support from the Independent Research Fund Denmark (Research grant 7027-00096B).
F. Calura acknowledges support from grant PRIN MIUR 2017 - 20173ML3WW\_001 and from the  INAF
Main-Stream (1.05.01.86.31).

\bibliographystyle{aa} 
\bibliography{disk}

\end{document}